\documentclass{aa}
\usepackage{graphicx}

\begin{document}

\newcommand{\Msolar}{\mbox{${\; {\rm M_{\sun}}}$}}
\newcommand{\Zsolar}{\mbox{${\; {\rm Z_{\sun}}}$}}
\newcommand{\ha}{\hbox{H$\alpha$}}
\newcommand{\oii}{\hbox{[O\,{\sc ii}]$\lambda$3727}}
\newcommand{\nii}{\hbox{[N\,{\sc ii}]}}
\newcommand{\hb}{\hbox{H$\beta$}}
\newcommand{\hg}{\hbox{H$\gamma$}}
\newcommand{\hd}{\hbox{H$\delta$}}
\newcommand{\hi}{\hbox{H\,{\sc i}}}
\newcommand{\hii}{\hbox{H\,{\sc ii}}}
\newcommand{\etal}{\hbox{et\thinspace al.\ }}
\newcommand{\oiiha}{\hbox{[O\,{\sc ii}]/H$\alpha$}}
\newcommand{\abswha}{\hbox{$EW_{{\rm H}\alpha}^{\rm abs}$}}
\newcommand{\abswhb}{\hbox{$EW_{{\rm H}\beta}^{\rm abs}$}}
\newcommand{\lra}{\hbox{$L_{1.4 {\rm}}$}}
\newcommand{\lha}{\hbox{$L_{{\rm H}\alpha}$}}
\newcommand{\lhb}{\hbox{$L_{{\rm H}\beta}$}}
\newcommand{\loii}{\hbox{$L_{\rm [O\,{\sc II}]}$}}
\newcommand{\loiii}{\hbox{$L_{\rm [O\,{\sc iii}]}$}}
\newcommand{\lnii}{\hbox{$L_{\rm [N\,{\sc ii}]}$}}
\newcommand{\lsii}{\hbox{$L_{\rm [S\,{\sc ii}]}$}}
\newcommand{\lfir}{\hbox{$L_{\rm FIR}$}}
\newcommand{\lir}{\hbox{$L_{\rm IR}$}}
\newcommand{\fra}{\hbox{$F_{\rm 1.4}$}}
\newcommand{\fha}{\hbox{$F_{{\rm H}\alpha}$}}
\newcommand{\fhb}{\hbox{$F_{{\rm H}\beta}$}}
\newcommand{\foii}{\hbox{$F_{\rm [O\,{\sc II}]}$}}
\newcommand{\ffir}{\hbox{$F_{\rm FIR}$}}
\newcommand{\fir}{\hbox{$F_{\rm IR}$}}
\newcommand{\mum}{\hbox{$\mu{\rm m}$}}
\newcommand{\sfr}{\hbox{${\rm SFR}$}}
\newcommand{\sfrha}{\hbox{${\rm SFR}_{{\rm H}\alpha}$}}
\newcommand{\sfroii}{\hbox{${\rm SFR}_{\rm [O{\sc II}]}$}}
\newcommand{\sfrfir}{\hbox{${\rm SFR}_{\rm FIR}$}}
\newcommand{\sfrir}{\hbox{${\rm SFR}_{\rm IR}$}}
\newcommand{\sfrra}{\hbox{${\rm SFR}_{\rm 1.4}$}}
\newcommand{\sfruv}{\hbox{${\rm SFR}_{\rm UV}$}}
\newcommand{\micron}{\hbox{$\mu$m}}

\titlerunning{Spectroscopic study of BCGs: IV SFR}
\authorrunning{X. Kong}

\title{Spectroscopic study of blue compact galaxies}
\subtitle{IV. Star formation rates and gas depletion timescales}
\author{X. Kong 
\inst{1,2,3}
}
\mail{xkong@optik.nao.ac.jp}

\institute{
Max Planck Institute for Astrophysics, Karl-Schwarzschild-Str.
1, D-85741 Garching, Germany
\and
Center for Astrophysics, University of Science and Technology
of China, 230026, P. R. China
\and
National Astronomical Observatory, 2-21-1 Osawa, Mitaka, Tokyo 
181-8588, Japan
}

\date{Received ; accepted}

\abstract{
This is the fourth paper in a series studying star formation rates, 
stellar components, metallicities, and star formation histories of 
a blue compact galaxy (BCG) sample. Using  \ha, \oii, infrared (IR), 
radio (1.4\,GHz) luminosities and  neutral hydrogen (\hi) gas 
masses, we estimated star formation rates and gas depletion 
timescales of 72 star-forming BCGs. The star formation rates of the 
BCGs in our sample span nearly four orders of magnitude, from 
approximately 10$^{-2}$ to $10^2$\,M$_\odot$\,yr$^{-1}$, with a 
median star formation rate of about 3\,M$_\odot$\,yr$^{-1}$. The 
typical gas depletion timescale of BCGs is about one billion years. 
Star formation could be sustained at the current level only on a 
timescale significantly lower than the age of the universe before 
their neutral gas reservoir is completely depleted. 

To assess the possible systematic differences among different star 
formation rate indicators, we compared the star formation rates 
derived from \ha, \oii, IR, and radio luminosities, and investigated 
the effects from underlying stellar absorption and dust extinction. 
We found that subtracting underlying stellar absorption is very 
important to calculate both dust extinction and star formation rate 
of galaxies. Otherwise, the intrinsic extinction will be 
overestimated, the star formation rates derived from \oii\ and \ha\ 
will be underestimated (if the underlying stellar absorption and 
the internal extinction were not corrected from the observed 
luminosity) or overestimated (if an overestimated internal 
extinction were used for extinction correction). 
After both the underlying stellar absorption and the dust extinction 
were corrected, a remarkably good correlation emerges among \ha, 
\oii, IR and radio star formation rate indicators. Finally, we find 
a good correlation between the measured star formation rate and the 
absolute blue magnitude, metallicity, interstellar extinction of 
BCGs. Our results indicate that faint, low-mass BCGs have lower star 
formation rates. 

}

\maketitle

\section{Introduction}

Understanding the star formation history of the universe is the 
primary goal of much current research in astronomy (Madau \etal 1996; 
Steidel \etal 1999). Star formation rate (SFR) is a crucial 
ingredient to understand the star formation history of galaxies at 
all redshifts, as well as the global evolution of the Universe. To 
obtain this understanding, a reliable estimate of the SFR in 
individual galaxies is required. Many calibrations of SFR depend 
on the luminosity measured at various wavelengths, including radio, 
infrared (IR), ultraviolet (UV), optical spectral lines (such as 
\ha, \oii) and continuum (Kennicutt 1998). 
Using these SFR indicators, a number of studies of the SFR were 
performed at different redshift regimes (such as Gallego \etal 1995; 
Pettini \etal 1998; Sullivan \etal 2001). Unfortunately, the 
agreement among these SFR indicators at different wavelengths is 
poor; the underlying reasons for the differences are not well 
understood (Charlot \etal 2002).  

Blue compact galaxies (BCGs) are characterized by their blue color, 
compact appearance, high gas content, strong nebular emission lines, 
and low chemical abundances (\cite{kunth00}).  
In recent years BCGs have attracted a great deal of interest 
and have become key in understanding fundamental 
astrophysical problems. Most of the work carried out has focused 
on a statistical analysis of BCG samples by means of surface 
photometry (Papaderos et al. 1996; Vitores \etal 1996; Doublier 
\etal 1997; \"{O}stlin \etal 1999; Cair{\' o}s \etal 2001; Gil de 
Paz \etal 2003). The measured broadband colors are strongly affected 
by interstellar extinction and gaseous emission; therefore, optical 
broadband photometry alone does not allow us to derive their 
physical properties accurately.

Spectroscopic information is required since spectra of galaxies 
contain a wealth of additional information concerning the physical 
properties of galaxies. The continuum and the absorption lines 
provide information concerning the stellar content, while the 
nebular emission lines provide a measure of the SFR and the 
interstellar metallicity. 
During the last three decades, a number of major objective-prism 
surveys for extragalactic emission-line objects have been carried 
out.  Examples include the University of Michigan (UM) survey 
(MacAlpine, Smith, \& Lewis 1977), the Universidad Complutense de 
Madrid (UCM) survey (Zamorano \etal 1994, 1996), the KPNO 
International Spectroscopic Survey (KISS) (Salzer \etal 2000). Some 
BCG candidates and other emission line objects have been discovered. 
Since the objective-prism spectra have very low spectral dispersion 
and small wavelength range, they cannot be used to investigate the 
physical properties of BCGs. 
Spectroscopic follow-up observations were carried out to study the 
characteristics of the survey BCG candidates (\cite{salzer89}; 
\cite{terlevich91}; \cite{gallego96}). The observations were 
obtained with different telescopes at different observatories with 
different equipment and instrument setups. The spectra from these 
observations have relatively low dispersions and low S/N ratios.
 
We have undertaken an extensive long-slit 
spectral observation of 200 BCGs since 1997. Spectra were obtained 
at the Cassegrain focus of the 2.16 m telescope of National 
Astronomical Observatory of China. A 300 line mm$^{-1}$ grating was 
used, giving a dispersion of 4.8 \AA\,pixel$^{-1}$. All 200 spectral 
observations were finished by February 2004, and 97 luminous BCG 
spectra were studied in this series. The observations have been 
performed with the same instruments; the reductions and analyses are 
homogeneous.

The blue color, high gas content and strong nebular emission-line 
spectrum of BCGs indicate that they have a high star formation 
activity.  
One of the key parameters in developing an understanding of the 
formation and evolution of BCGs is the determination of the SFRs. 
While SFRs are central to discussions of the evolution of BCGs, there 
are surprisingly few quantitative SFRs for BCGs available in the 
literature.  In addition, the values available in the literature 
suggest that the SFRs of BCGs might vary over five orders of 
magnitude, and the mean SFRs given in the existing investigations 
are noted to also vary over two orders of magnitude. It is therefore 
not clear whether and to what extent the wide range of BCG SFRs 
results from the use of different techniques, assumptions, or sample 
selections, employed in the available studies. 

Furthermore, the studies of intermediate redshift compact galaxies 
in the Hubble deep field (\cite{guzman96}; \cite{phillips97}) 
reveal that these include galaxies with properties very similar to 
those of local luminous BCGs, such as small sizes, high luminosities 
and blue colors. 
Because spectroscopic observation for these intermediate redshift 
galaxies is difficult (for example, \ha\ is shifted outside the 
optical range), some other SFR indicators or methods must be used 
to determine the SFRs of these intermediate redshift galaxies 
(\cite{guzman97}; \cite{hammer01}; \cite{guzman03}). Therefore, 
estimating SFRs of nearby BCGs with different SFR indicators may 
help us in accurately estimating SFRs of intermediate redshift 
galaxies and studying their evolutionary history. 

Based on our homogeneous BCG optical spectral sample, we calculate 
SFRs and gas depletion timescales of a sample of star-forming blue 
compact galaxies, using the \ha\ and \oii\ emission line fluxes and 
the IR and radio luminosities. We compare the global SFRs derived 
from \oii\ and \ha, with or without corrections for dust extinction 
and underlying stellar absorption, to the SFRs derived from the IR 
and radio luminosities of these galaxies and investigate the source 
of the discrepancies among these different SFR indicators. We also 
investigate the relationship between the intrinsic SFR and other 
galaxy properties. We found that subtracting underlying stellar 
Balmer absorptions is very important in estimating dust extinction 
and SFR of galaxies. 

The paper is organized as follows. In Sect. 2 we review the 
spectroscopic observation and reduction, and describe the {\it IRAS} 
data, the radio 1.4\,GHz observations. In Sect. 3 we calculate 
multiwavelength SFR estimates for the galaxy sample using the 
available flux data. In Sect. 4, we compare the results obtained 
using the different SFR estimators and investigate the underlying 
reason for their discrepancies. We present some discussion in Sect. 
5 and our conclusions in Sect. 6. 
Throughout this paper we assume 
$H_0=71\,\rm{km\,s^{-1}\,Mpc^{-1}}$, and adopt the Salpeter 
initial mass function with mass ranges $0.1\,\Msolar < {\rm M} < 
100\,\Msolar$.

\section{The data}

\subsection{{\rm \ha\ }and {\rm \oii\ } emission line fluxes}

To investigate the current SFRs, stellar components, metallicities 
and star formation histories of BCGs, we have observed an atlas of 
optical spectra of the central regions of 97 blue compact galaxies 
in the first paper of this series (\cite{kong02a}). Since we want 
to combine the optical spectra we obtained with \hi\ data to 
constrain simultaneously the stellar and gas contents of BCGs, our 
sample galaxies were selected from large \hi\ surveys (Gordon 
\& Gottesman 1981; Thuan \& Martin 1981). The spectra were obtained 
at the 2.16 m telescope at the XingLong Station of the National 
Astronomical Observatory of China. A 300 line mm$^{-1}$ grating was 
used to achieve coverage in the wavelength region from 3580 to 7400 
\AA\ with about 4.8\,\AA\ per pixel resolution. The emission line 
equivalent widths and fluxes for our BCG sample were provided in 
the second paper of this series (\cite{kong02b}). The typical 
uncertainties of the measurements are less than 10\% for \ha6563, 
\hb4861, and \oii\, emission lines.  

\subsubsection{Stellar absorption and extinction correction}

\begin{figure*}
\centering
\includegraphics[angle=-90,width=\textwidth]{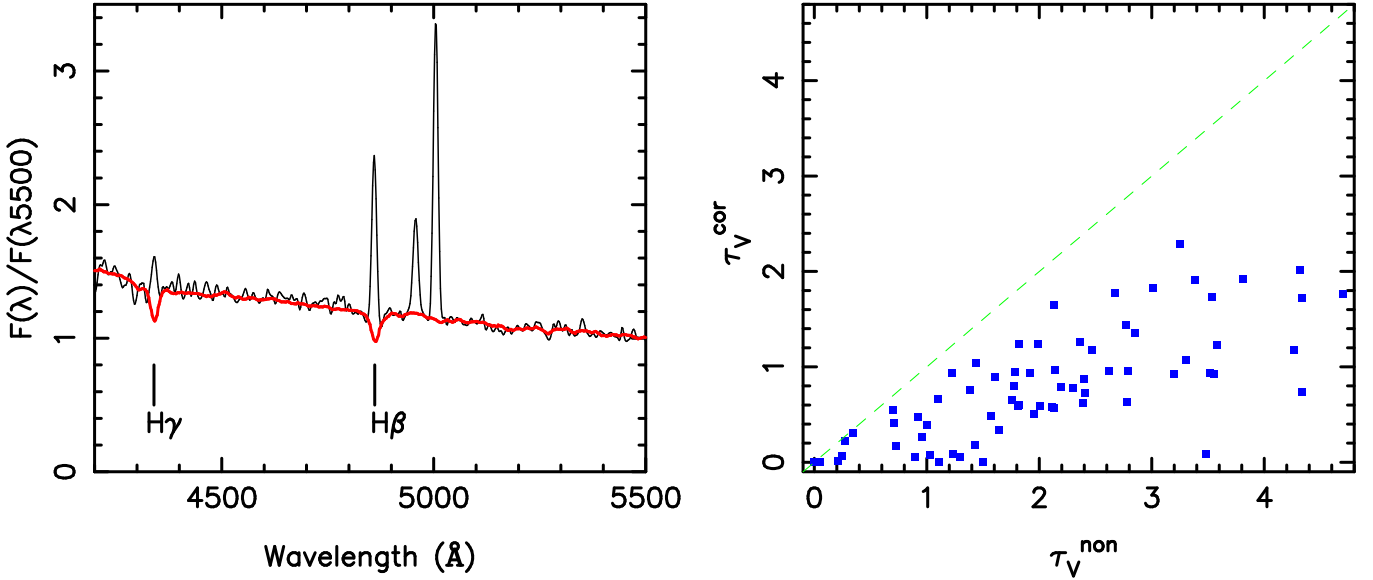}
\caption{
{\it Left:} 
Rest-frame spectrum of Haro 35 on which a synthetic spectrum 
({\bf thick line}) from the empirical population synthesis 
method. 
{\it Right:} 
Effect of underlying stellar absorption to dust extinction 
$\tau_V$. $\tau_V^{cor}$ and $\tau_V^{non}$ were calculated 
from the fluxes \ha\ and \hb, with and without corrections 
for the underlying stellar absorption. The dashed line 
presents $\tau_V^{non}$ = $\tau_V^{cor}$. 
\label{ebv}}
\end{figure*}

The underlying stellar absorption affects accurate measurement of 
H-Balmer emission line, and is crucial to constrain the 
attenuation by dust and the SFRs of galaxies.  Some previous 
works have discussed this effect on extinction and SFR estimation.
Rosa-Gonz{\' a}lez \etal (2002) found the \sfrha\ is close to the 
\sfrir\ while both \sfroii\ and \sfruv\ show a clear excess for 31 
nearby star-forming galaxies. The underlying Balmer absorption 
results in an underestimate of the emitted fluxes and an 
overestimation of the internal extinction of the galaxy. 
Subtracting the underlying absorption from emission lines (with 
EW$_{abs}$(\ha) = EW$_{abs}$(\hb) = EW$_{abs}$(\hg)) brings into 
agreement the SFR in the optical with those in the FIR. 
Using 149 galaxies from the Stromlo-APM redshift survey and using
Charlot \& Longhetti (2001) model (which provides a physically 
consistent description of the effects of stars, gas and dust on the 
integrated spectra of galaxies), Charlot \etal (2002) found the 
contamination of Balmer emission by stellar H-Balmer absorption is 
important in estimating SFRs of galaxies with small observed 
\ha\ emission equivalent widths.
Flores \etal (2004) studied SFRs for a sample of 16 distant galaxies; 
the underlying Balmer absorption lines were estimated by the Bica \& 
Allion (1986) method. After subtracting the underlying Balmer 
absorption line effect, the extinction estimates using the \hb/\hg\ 
and the \ha/\hb\, and the SFR estimates (for galaxies with \sfrir\ 
below $\sim$100 M$_\odot$\,yr$^{-1}$) using \sfrha\ and \sfrir\ are 
in excellent agreement.

To derive the equivalent widths for these underlying stellar 
absorption lines and correct the measured Balmer emission line 
fluxes for these absorptions, we have applied an empirical 
population synthesis method to our BCG spectra (\cite{cid01}). 
The left panel of Figure~\ref{ebv} illustrates the spectral fit of 
one of our sample galaxies. The synthetic spectrum ({\bf thick line}) 
from an empirical population synthesis analysis provides good fits to 
the observed spectrum of BCGs (thin line). The absorption wings of 
\hb\ and \hg\ in the observed spectrum is well reproduced by the 
synthetic spectrum too (\cite{kong03}).
The synthetic stellar spectrum is then subtracted from the observed 
spectrum to yield a pure emission spectrum uncontaminated by 
underlying stellar absorption, which we can use to measure the 
Balmer decrement, derive extinction and estimate SFR.

In the right panel of Figure~\ref{ebv}, we show the effect of 
underlying stellar absorption on the internal extinction of galaxies 
($\tau_V = 0.921 A_V$ is the effective absorption optical depth at 
5500\AA\ by interstellar dust), where $\tau_V$ was determined 
using these two strongest Balmer lines, \ha, \hb, and the Charlot 
\& Fall (2000) effective absorption curve $\tau_{\lambda} = 
\tau_V(\lambda/5500{\rm \AA})^{-0.7}$ 
(which is similar to the Calzetti (2001) effective absorption curve at 
optical wavelengths; $k(\ha)-k(\hb)=-0.57$ for the Charlot \& Fall (2000) 
effective absorption curve is almost same as $k(\ha)-k(\hb)=-0.58$ 
for the Calzetti (2001) effective absorption curve and varying the 
extinction curves has negligible effects in the visible range).
$\tau_V^{non}$ was calculated from the \fha\ and \fhb, where the 
underlying stellar absorption was not subtracted.  $\tau_V^{cor}$ 
was calculated from the \fha\ and \fhb, and the underlying stellar 
population absorption was subtracted. From this figure, we find 
that subtracting underlying stellar absorption is very important 
in calculating the $\tau_V$; otherwise the internal extinction of 
galaxy will be overestimated. Using $\tau_V^{cor}$, the internal 
extinction of each galaxy was corrected.

\subsubsection{Aperture corrections}

In addition to the underlying absorption and extinction correction, 
the emission line fluxes also require an aperture correction to 
account for the fact that only a limited amount of emission from 
a galaxy is detected using a slit 2\arcsec$\sim$ 3\arcsec\ wide 
and 3\arcmin\ long. 

Two different methods were used to estimate the slit covering 
fraction. The first method is similar to that used by Kewley \etal\ 
(2002). We convolve the galaxy spectral energy distribution with 
the normalized B-filter response curve (Bessell 1979) in $B$-band 
magnitude, $m_B(spec)$, and estimate the fraction, $f$, by 
comparison with the B-band photometry for the whole galaxy: $f = 
10^{-0.4 [m_B-m_B(spec)]}$. $m_B$ is the B-band magnitude for the 
whole galaxy, which is taken from Gordon \& Gottesman (1981), de 
Vaucouleurs \etal (1991, RC3) and the NASA/IPAC Extragalactic 
Database (NED).

The second method is similar to that used by Carter \etal\ (2001). 
We use the digitized Palomar (POSS) images of BCGs to estimate the 
slit covering fraction for the galaxies. We measure the ratio of 
the photographic flux within the slit aperture to the total 
photographic flux, $f = F_{slit}/F_{total}$, where $F_{slit}$ is 
the photographic flux within the slit aperture, and $F_{total}$ is 
the total photographic flux of a galaxy.
 
The results from these two different methods are in good agreement 
in most cases. We use the average slit covering fraction from these 
two methods for the aperture correction. Most of our spectra contain 
20$-$50\% of the total flux of the galaxies; for the largest galaxies 
this fraction can be as small as 10\%; for the smallest galaxies 
this fraction can be as large as 90\%. The average slit fraction 
of our sample is 37\%. The absolute fluxes of \ha\ and \oii, 
$F_{abs}$, were calibrated from the spectral flux, $F_{spec}$, and 
the average slit covering fraction with $F_{abs} = F_{spec}/f$.

\subsection{Infrared data}

Indicators of SFR at longer wavelengths are less effected by dust 
extinction than those at UV or optical wavelengths. To investigate 
the underlying reasons for the discrepancies between different SFR 
indicators, we also use the infrared and radio luminosities to 
derive the SFR of BCGs.

{\it Infrared Astronomical Satellite} ({\it IRAS}) products include 
12, 25, 60 and 100~\mum\ flux densities for galaxies from the Point 
Source Catalog (brighter than about 0.5\,Jy) and Faint Source 
Catalog (typically brighter than 0.2\,Jy) (Moshir \etal 1992). We 
cross-correlated our galaxies with the {\it IRAS} Faint Source 
Catalogue and Point Source Catalogue. The {\it IRAS} beam size at 
60~\mum\ is 1.5\arcmin, and the {\it IRAS} positional uncertainty 
is 30\arcsec. Because the spectroscopic positional uncertainty is 
only a few arcseconds and because all BCGs are smaller than 2\arcmin\ 
in extent, we used a detection radius of 30\arcsec. Any {\it IRAS} 
source associated with a BCG should be detected within this radius. 
We found that of the 72 BCGs in our sample, 58 are detected at all 
four 
bands by {\it IRAS} within 30\arcsec; most of them have moderate 
or good quality fluxes at 60 and 100~\mum.

For each galaxy that was detected by {\it IRAS}, the total 
far-infrared flux $F_{\rm IR} = F(8-1000~\mum)$ was estimated from 
the observed {\it IRAS} flux densities at 25, 60, and 100~\mum, using 
the method of Dale \& Helou (2002). The total IR flux 
determined in this way is typically 2.2 $\pm$ 0.4 times larger than 
the {\it FIR} estimator of Helou \etal~(1985), and similar to that 
from the estimator of Sanders \& Mirabel (1996).

\subsection{1.4 GHz radio flux}

The insensitivity of radio wavelengths to dust obscuration makes 
radio emission a particularly attractive way of estimating SFRs in 
star-forming galaxies. To calculate the SFRs of BCGs, we extract 
the radio 1.4 GHz integrated flux for our sample galaxies from the 
NRAO VLA Sky Survey ({\it NVSS}) and Faint Images of the Radio Sky 
at Twenty-cm ({\it FIRST}) survey.

The {\it NVSS} is a 1.4 GHz continuum survey covering the entire 
sky north of -40 deg declination, made using the Very Large Array 
(VLA) operated by the National Radio Astronomy Observatory. About 
220,000 individual snapshots (phase centers) have been observed.  
The resulting images have a rms noise level of about 0.5\,mJy\, 
beam$^{-1}$ and a FWHM resolution of 45\arcsec. A detailed 
description is given in Condon \etal (1998). For 45 galaxies in our 
BCG sample, an {\it NVSS} catalog source is close enough ($<$ 
30\arcsec) to be a probable identification.

The {\it FIRST} is a 1.4\,GHz radio continuum survey currently 
covering about 9033 square degrees of the North Galactic Cap. The 
images have a typical rms noise level of 0.15\,mJy with a FWHM 
resolution of 5\arcsec\ (White \etal 1997). From the {\it FIRST} 
catalog (03apr11 version, 2003 April 11), 24 galaxies were matched 
with our BCG sample ($< 30''$). For these 24 {\it FIRST} galaxies, 
22 of them were observed by the {\it NVSS} too. Because of the much 
higher resolution of {\it FIRST} compared to {\it NVSS}, we will 
use the {\it FIRST} flux density for SFR calculation when galaxies 
were observed by both {\it FIRST} and {\it NVSS}.  
\subsection{Distances}

Of the 72 star-forming BCGs in our sample, all have \oii\ and \ha\ 
fluxes, 58 have IR fluxes, and 47 have 1.4 GHz radio fluxes.  To 
calculate the SFRs of galaxies, we convert the fluxes into 
luminosities, using $L = 4 \pi D^2 \times F$. Here $D$ is the 
cosmological luminosity distance, calculated using the 
heliocentric velocities of the galaxies and a Hubble constant value 
of $H_0 = 71$\,km\,s$^{-1}$\,Mpc$^{-1}$ from WMAP (Spergel \etal 
2003).

\section{Multiwavelength star formation rates}

In this section, we calculate the SFRs of 72 BCGs from the 
luminosities of \oii\, and \ha\ nebular lines, IR and radio 
continuum using well-known calibrations.

\subsection{SFR from \ha\ luminosities}

Hydrogen recombination line fluxes have been used very extensively 
to estimate the SFR, since they are proportional to the number of 
photons produced by the hot stars, which is in turn proportional 
to their birthrate.  Most applications of this method have been 
based on measurements of the \ha\ line. Other hydrogen recombination 
lines, such as \hb\ and \hg, were not widely used because the effect 
of the underlying stellar absorption and dust extinction are more 
important in these blue lines.

For the estimate of the SFR from \ha\ luminosities we used the 
expression given by Kennicutt (1998):

\begin {equation} 
\sfrha\,( \rm{M}_{\odot}\,\rm{yr}^{-1}) =
7.9\times 10^{-42} \lha\,(\rm{erg\,s}^{-1}).
\label{eqha}
\end {equation}

\noindent which is valid for a T$_e$=10$^4$K and Case B 
recombination, i.e. all the ionizing photons are processed by the 
gas. 

Figure 2a shows that the SFRs derived from \ha\ fluxes for the 72 
star-forming blue compact galaxies in our sample. \sfrha\ of BCGs 
cover almost four orders of magnitude, from 0.01 to 65 
$\rm{M}_{\odot}\,\rm{yr}^{-1}$ and with a median SFR of 2.4 
$\rm{M}_{\odot}\,\rm{yr}^{-1}$.

\begin{figure}
\centering
\includegraphics[angle=-90,width=0.45\textwidth]{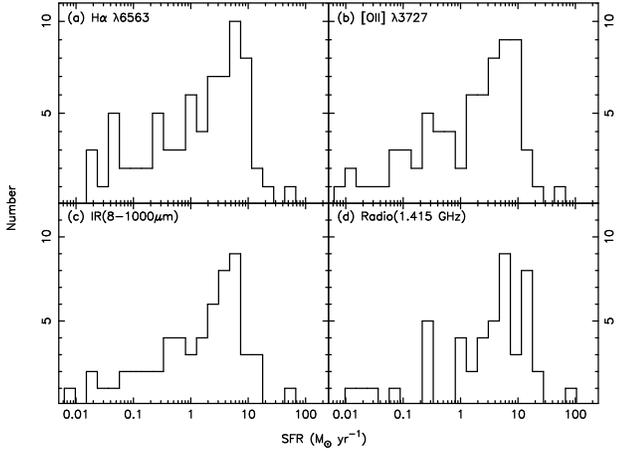}
\caption{
Histograms showing the range of derived SFRs for the star-forming 
BCGs. 
It indicates the broad range of SFRs exhibited in blue compact 
galaxies, 
from approximately 10$^{-2}$ to several times 
10\,M$_\odot$\,yr$^{-1}$. 
The median SFRs from \oii, \ha, IR and radio luminosities are 2.8, 
2.4, 
3.6 and 5.3\,M$_\odot$\,yr$^{-1}$, respectively.
\label{sfrhist}}
\end{figure}

\subsection{SFR from \oii\ luminosities}

If recombination lines are unavailable, the collisionally excited 
lines of heavy elements may be used in their stead. The \oii\, 
doublet is the most natural star formation tracer for objects with 
redshifts larger than 0.4 where \ha\ is shifted outside the optical 
range. The \oii\ luminosity is not directly coupled to the 
ionization rate, and their excitation is known to be sensitive to 
the abundance and the ionization state of the gas. However, it is 
claimed that the excitation of \oii\ is sufficiently stable in 
observed galaxies that it can be calibrated {\em empirically} with 
\ha\ as a quantitative SFR tracer (Gallagher \etal 1989; Kennicutt 
1992). 

Figure~8a in Kong \etal(2002) shows a plot of the \oii\ emission 
line flux versus the \ha\ emission line flux corrected for stellar 
absorption in our star-forming BCG sample. There is a good 
correlation between \foii\ and \fha. From the data in Fig.~8a of 
Kong \etal (2002), we derive a least-squares relationship of 
\foii/\fha = 0.85. The \foii/\fha\ of BCGs (local galaxies) in good 
agreement with the values of high redshift galaxies, where 
\foii/\fha $\sim$ 0.9 (Hippelein \etal 2003).  
Adopting the calibration between SFR and \ha\ luminosity given in 
Kennicutt (1998), the SFR and \loii\ can be expressed as:

\begin {equation}
\sfroii\,( \rm{M}_{\odot}\,\rm{yr}^{-1}) =9.3 \times 10^{-42} 
\loii\, (\rm{erg\,s}^{-1}).
\label{eqoii}
\end {equation}

This formula is similar to that given by Gallagher \etal (1989) using 
a sample of 75 blue galaxies, where $\sfroii\,(\rm{M}_{\odot} 
\rm{yr}^{-1}) =1.0 \times 10^{-41} \loii\, (\rm{erg\,s}^{-1})$ for 
blue galaxies. Using the Eq.~(\ref{eqoii}), we calculate the 
\sfroii\ for 72 star-forming BCGs in our sample, and show the results 
in Figure\ref{sfrhist}b. The range of \sfroii\ from 0.01 to 67 
$\rm{M}_{\odot}\,\rm{yr}^{-1}$ and with a median SFR of 2.8 
$\rm{M}_{\odot}\,\rm{yr}^{-1}$.

\subsection{SFR from far-infrared continuum}

A significant fraction of the bolometric luminosity of a galaxy is 
absorbed by interstellar dust and re-emitted in the thermal IR, at 
wavelengths of roughly 10--300~\mum. The absorption cross section 
of the dust is strongly peaked in the ultraviolet, so in principle 
the FIR emission can be a sensitive tracer of the young stellar 
population and SFR. Using theoretical stellar flux distributions 
and evolutionary models, we can then derive the SFRs of galaxies. 

The SFR of a galaxy in our sample was estimated from its integrated 
IR luminosity (Kennicutt 1998) by: 

\begin {equation}
\sfrir\,(\rm{M}_{\odot}\,\rm{yr}^{-1}) =4.5 \times 10^{-44}
\lfir\,({\rm erg\,s}^{-1}),
\label{eqfir}
\end {equation}

\noindent where $L_{\rm{FIR}}$ refers to the luminosity integrated 
over the full mid-infrared to submillimeter spectrum (8--1000 \mum).  
The \sfrir\ for 58 BCGs in our sample with {\it IRAS} detection were 
plotted in Figure~\ref{sfrhist}c, range from 0.02 to 67 
$\rm{M}_{\odot}\, \rm{yr}^{-1}$ and with a median SFR of 3.6 
$\rm{M}_{\odot}\, \rm{yr}^{-1}$.

\subsection{SFR from 1.4\,GHz continuum}

In theory, the radio emission results predominantly from the 
nonthermal supernova remnant flux, and therefore should be very 
model dependent and critically influenced by the density of, and 
magnetic field in, the interstellar gas. However, the FIR/radio (the 
1.4 GHz radio continuum flux) correlation suggests that the constant 
of proportionality is nearly the same for all star forming galaxies 
and is insensitive to unknown variables such as magnetic field 
strength (Condon 1992; Yun \etal 2001). 
The reality of this correlation has been repeatedly demonstrated 
by observation, not only in terms of the integrated flux, but also 
in terms of spatial correlations between these two quantities within 
individual galaxies. Therefore, the radio continuum flux can be used 
as a star formation indicator. 

For an individual star-forming galaxy, the SFR is directly 
proportional to its radio luminosity at 1.4\,GHz (Condon 1992):

\begin{equation}
\sfrra\,( \rm{M}_{\odot}\,\rm{yr}^{-1}) = 
{\it n} \times 2.2 \times 10^{-22} \lra\, (\rm{W\,Hz}^{-1}).
\label{eqora}
\end {equation}

Condon (1992) derives this relation by calculating the synchrotron 
radio emission from supernova remnants and the thermal radio 
emission from HII regions.  Both the thermal and non-thermal 
components of the radio expression are proportional to the formation 
rate of high-mass stars ($M > 5 {\rm M}_{\odot}$) which produce 
supernova and large HII regions, so the factor $n$ is included to 
account for the mass of {\it all} stars in the interval $0.1-100{\rm 
M}_{\odot}$. We have assumed throughout a Salpeter IMF ($x=2.35$), 
for which $n=5.5$.  

The \sfrra\ for 47 BCGs in our sample with NVSS and FIRST detection 
were plotted in Figure~\ref{sfrhist}d, range from 0.01 to 129 
$\rm{M}_{\odot}\, \rm{yr}^{-1}$ and with a median SFR of 5.3 
$\rm{M}_{\odot}\, \rm{yr}^{-1}$. Most of them are consistent with 
those derived from \ha, \oii\ and IR luminosities, but the \sfrra\ 
of I Zw 56 is much larger than those derived from other SFR 
indicators. 

\subsection{\sfr\ of BCGs}

Based on the FIR (60\,$\mu$m) and radio continuum (1.4\,GHz) 
luminosities, Izotova \etal (2000) derived the SFRs of 27 BCGs, with 
a mean of about 18\,M$_\odot$\,yr$^{-1}$.  Using spectroscopically 
derived H$_{\beta}$ luminosities, Popescu \etal (1999) determined 
the SFRs of a sample of BCGs, with a mean SFR of about 
0.5\,M$_\odot$\,yr$^{-1}$.  Therefore, the values available in the 
literature suggest that the mean SFRs given in the existing 
investigations vary over two orders of magnitude.  It 
is therefore not clear whether and to what extent the wide range 
of BCG SFRs results from the use of different techniques, or sample 
selections, employed in the available studies.

In this section, we have determined the SFRs of 72 blue compact 
galaxies, using the \ha, \oii\, emission line fluxes, the {\it IRAS}, 
and the 1.4\,GHz radio continuum luminosities.  Since we have 
determined the SFRs for our homogeneous BCG sample, and calculated 
the median SFR of the sample using these different SFR indicators, 
it can be used to understand the reasons for these discrepancies 
in the literature.  

Fig.~\ref{sfrhist} shows that the SFRs derived from different star 
formation indicators cover almost four orders of magnitude, from 
$10^{-2}$ to $10^{2}$ M$_\odot$\,yr$^{-1}$.  After correcting for the 
underlying stellar absorption and the dust extinction effects, the 
SFR of each galaxy derived from those different SFR indicators 
is similar. Using different techniques will not result a large 
different median SFR of galaxies if we have observed the same sample 
galaxies over the entire wavelength bands. 

From Fig.~\ref{sfrhist}, we also found that the medium and average 
SFRs from these different indicators of our sample galaxies are 
different; values derived from the IR and radio data are much 
larger than those derived from the optical emission lines. Therefore, 
the sample selection effect is an important reason for the different 
mean SFR of BCGs in the literature. This can be explained by: the 
positional accuracy of far-infrared observation being not good, 
therefore, it is difficult to identify an object with weak IR 
emission. On the opposite side, radio observations typically have 
better positional accuracy, but the detector in the radio band is not 
sensitive. Therefore both far-infrared and radio observations are 
difficult to use to identify an objects with weak star formation activity; 
these survey are biased towards the galaxies with high SFRs, and the mean 
values of these survey samples are larger than those of optical 
observations.

\section{Comparison of SFRs derived using different standard 
estimators}

We have applied the commonly used SFR estimators to the 72 
star-forming BCGs in our sample. In this section, we will 
investigate the SFR relation from these different standard 
estimators and the underlying reason for their discrepancies.

\subsection{\sfroii\ and \sfrha}

\begin{figure}
\centering
\includegraphics[width=0.45\textwidth]{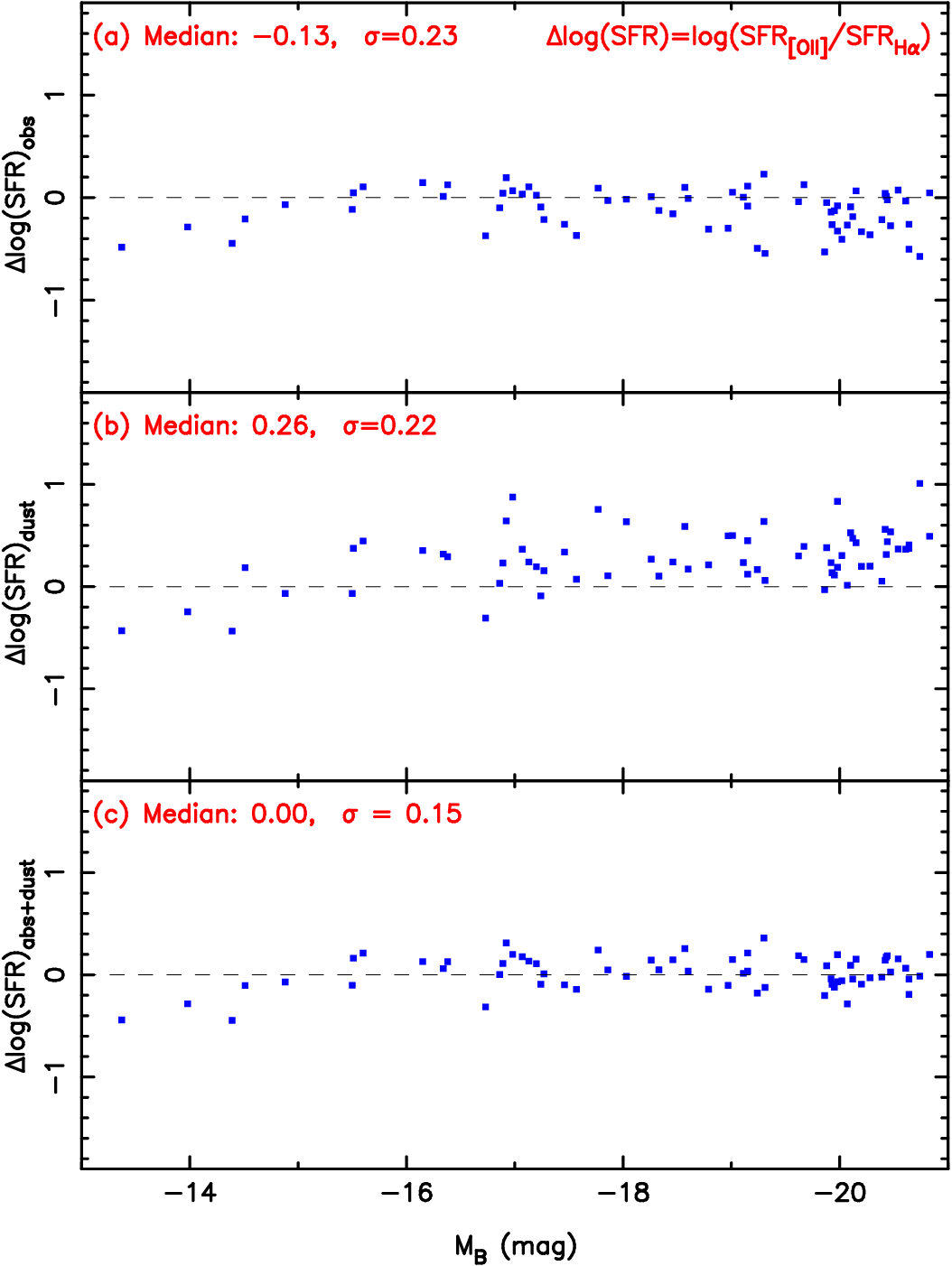}
\caption{
Logarithmic difference between \sfrha\ and \sfroii\ plotted 
against absolute rest-frame ${\rm M}_{\rm B}$ magnitude for 
star-forming BCGs. 
SFRs derived from \lha\ and \loii: 
(a) before applying any correction for \lha\ and \loii; 
(b) after correction for dust extinction, using the 
$\tau_V^{non}$ (see Fig.1); 
(c) after further correction for the underlying stellar 
absorption.
In each panel, the median $\Delta\log(\sfr)$ offset is 
indicated along with the associated mean offset and the rms 
scatter.
\label{oii-ha}}
\end{figure}

Figure~\ref{oii-ha} shows the logarithmic difference between 
\sfrha\ and \sfroii\ for the 72 BCG in our sample as a function of 
the B-band absolute magnitude, ${\rm M}_{\rm B}$. In each panel, 
we indicate the median $\Delta\log(\sfr)$ offset and the rms 
scatter.

In Fig.~\ref{oii-ha}a, \sfroii\ and \sfrha\ were calculated with 
equations~(1), (2) and the observed \loii\ and \lha\ data, where 
the underlying stellar absorption and the internal extinction were 
not corrected. Fig.~\ref{oii-ha}a is consistent with the recent 
result by Charlot et al. (2002) that bright galaxies tend to have 
lower \oiiha\ ratios than less luminous galaxies. We find that, for 
${\rm M}_{\rm B}<-18$ mag, \oii-derived SFR estimates are typically 
30 per cent smaller than \ha-derived ones, while at fainter 
magnitudes, the two estimators give similar results. At fixed 
magnitude, the rms scatter in \sfroii/\sfrha\ is a factor of $\sim 
1.7$. The scatter in this plot and the weak trend with magnitude 
can probably be explained by variations in the effective gas 
parameters (ionization, metallicity, dust content) of the galaxies. 

In Fig.~\ref{oii-ha}b, SFRs were calculated from the \loii\ and 
\lha\ data for which the internal extinction was corrected by 
$\tau_V^{non}$, which was determined by the observed \ha$/$\hb\ of 
the galaxy (the underlying stellar absorption was not subtracted). 
In star-forming galaxies, the Balmer emission lines from the ionized 
gas appear superimposed on the stellar absorption lines; this effect 
grows in importance towards the higher order Balmer lines, and the 
result is an underestimate of the emitted fluxes and an 
overestimate of the internal extinction, as show in Fig. 1. This 
overestimated internal extinction correction makes the \oii-derived 
SFR estimates typically a factor of $\sim 2$ higher than 
\ha-derived ones.  

In Fig.~\ref{oii-ha}c, the \sfroii\ and \sfrha\ were determined by 
the \loii\ and \lha, where the underlying stellar absorption was 
subtracted and the internal extinction was corrected. The SFR of 
BCGs from \foii\ and \fha\ have an excellent match, which can be 
understood easily since we derived the \sfroii\ formula (equation 
2) from the relation between \foii\ and \fha. 
In Fig.~\ref{oii-ha}c the rms scatter in \sfroii/\sfrha\ becomes 
smaller, about 1.4. 
The 3 low luminosity BCGs are unaffected by stellar absorption and 
extinction. This is because: 1) They are low metallicity galaxies, 
and have a small dust component. The effect of internal extinction 
is not important for them. 2) They appear dominated by a recent 
burst of star formation, which causes their extremely strong 
emission line spectrum. The underlying stellar absorption is much 
weaker than the emission component. 
In addition, the correlation between \fha\ and \foii\ 
derived from the whole sample (most of them are high luminosity 
galaxies) is not good for these low metallicity galaxies, and may 
underestimate the \sfroii\ for low luminosity galaxies. 

\subsection{\sfrir\ and \sfrha}

\begin{figure}
\centering
\includegraphics[width=0.45\textwidth]{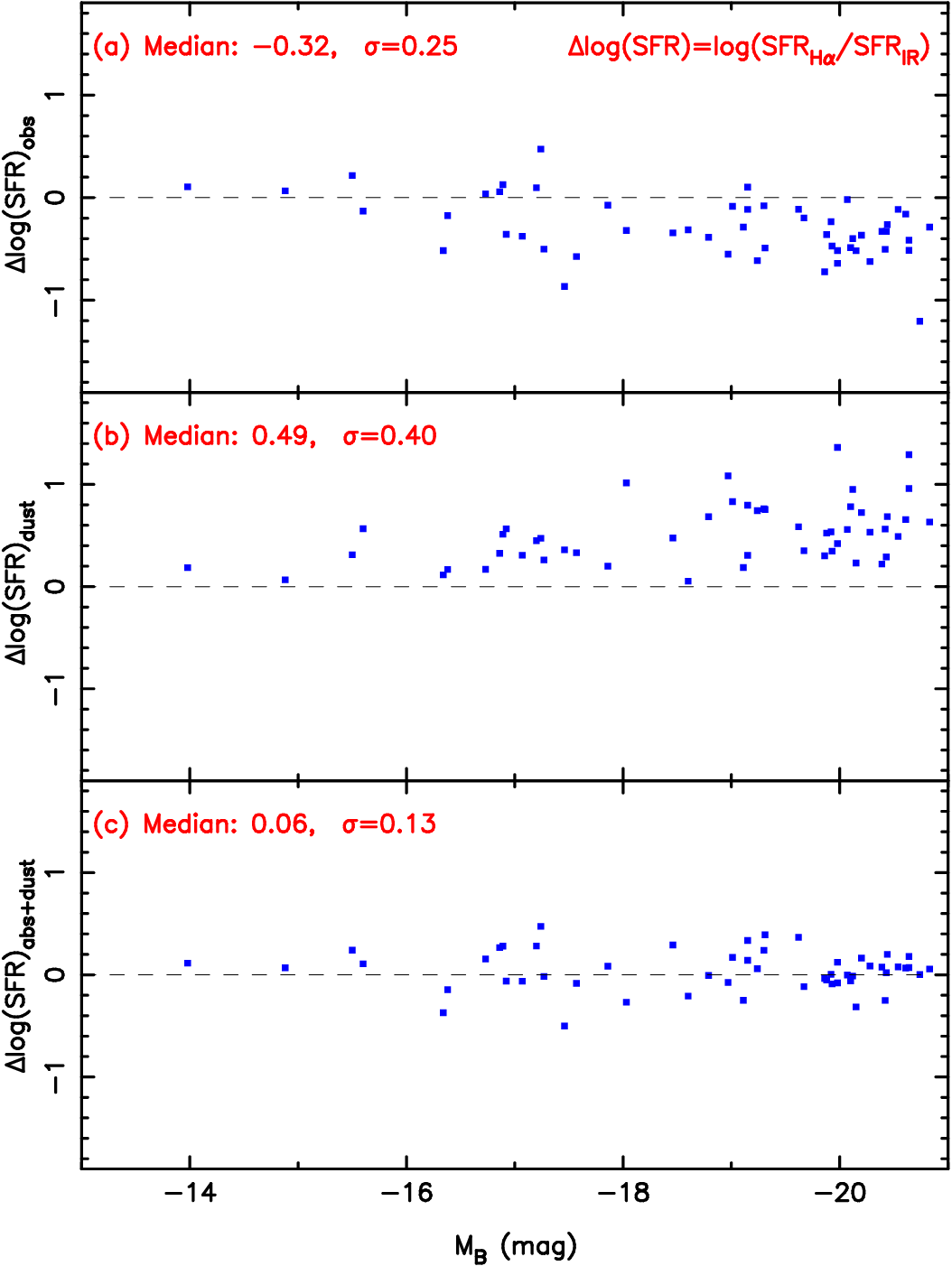}
\caption{
Logarithmic difference between \sfrha\ and \sfrir\ plotted against 
absolute rest-frame ${\rm M}_{\rm B}$ magnitude for 58 BCGs, which
were observed by {\it IRAS}. 
SFRs derived from \lha\ and \lfir: 
(a) before applying any correction for \lha\, 
(b) after correction for dust extinction, using the $\tau_V^{non}$, 
(c) after further correction for the underlying stellar absorption. 
In each panel, the median $\Delta\log(\sfr)$ offset is indicated 
along with the associated mean offset and the rms scatter.
\label{ha-ir}}
\end{figure}

Figure~\ref{ha-ir} shows the logarithmic difference between 
\sfrha\ and \sfrir\ for 58 BCGs ({\it IRAS} data are available) in 
our sample as a function of ${\rm M}_{\rm B}$.  

In Fig.~\ref{ha-ir}a, \sfrir\ was derived from equation~(3), 
\sfrha\ was calculated with equation~(1) and the observed \ha\ 
luminosity data, \lha. The underlying stellar absorption and the 
internal extinction were not corrected in the observed \ha\ 
luminosity data in this panel. We find the difference between the 
SFRs estimated using \lha\ and those estimated using the total FIR 
luminosity is large.  
The SFRs derived using \ha\ are lower than those obtained using \fir. 
The median discrepancy tends to increase with B-band absolute 
luminosity and amounts to a factor of 2 at ${\rm M}_{\rm B} < -18$. 
The discrepancy and the scatter is similar to the SFRs derived using 
\oii. The median and the standard deviation of the logarithmic 
difference between \sfrha\ and \sfrir\ for 58 BCGs in our sample 
is $-0.32 \pm 0.25$, which similar to (or a little smaller than) 
that of Charlot \etal (2002) ($-0.47\pm0.38$) and of Rosa-Gonz{\' 
a}lez \etal (2002) ($-0.53\pm0.52$).

In Fig.~\ref{ha-ir}b, \sfrha was calculated from the \lha\ data with 
the internal extinction corrected by $\tau_V^{non}$ (see 
Fig.~1 for more detail). Since the effect of the underlying stellar 
absorption was not considered, the median of the $\tau_V^{non}$ is 
about 1.9, which is much larger than the median of the intrinsic 
internal extinction value ($\tau_V^{cor}\sim0.8$) of galaxies.
The internal extinction value of galaxies was overestimated; it 
makes the differences between the SFRs estimated using optical lines 
and those estimated using the total IR luminosities greater. 
Fig.~\ref{ha-ir}b shows that \ha-derived SFR estimates are 
typically a factor of $\sim 3$ higher than \fir-derived ones.

For the galaxies in our sample, the mean of the equivalent width 
of H-Balmer absorption corresponds to $\abswha\approx -2$ \AA\, 
$\abswhb\approx -4$ \AA\ (Kong \etal 2003), which is consistent with 
the results from Bruzual \& Charlot(2003), $\abswha\ \approx 0.6 
\abswhb$. The correction for stellar absorption is therefore quite 
important for galaxies with narrow observed \ha\ emission equivalent 
widths. This is shown in Fig.~\ref{ha-ir}c, where we corrected the 
\ha-derived SFRs of individual galaxies in our sample for both the 
underlying stellar absorption and the interstellar dust extinction 
(using $\tau_V^{cor}$). These corrections bring the \ha-derived 
SFRs into agreement with the \ffir-derived SFRs.  

The median and the standard deviation of the logarithmic difference 
between \sfrha\ and \sfrir\ for 58 BCGs in our sample is $0.06 \pm 
0.13$. Our results agree with those of Rosa-Gonz{\' a}lez 
\etal (2002), and also Charlot \etal (2002).  
Rosa-Gonz{\' a}lez \etal (2002) developed a method to 
estimate simultaneously the intrinsic visual extinction $A_V^*$ and 
the underlying Balmer absorption. The values of $A_V^*$ were applied 
to the emission line fluxes. They also found that the extinction 
correction which includes the effects of an underlying stellar 
Balmer absorption brings into agreement \sfrha\ and \sfrir.

\subsection{\sfrra\ and \sfrha}

\begin{figure}
\centering
\includegraphics[angle=-90,width=0.45\textwidth]{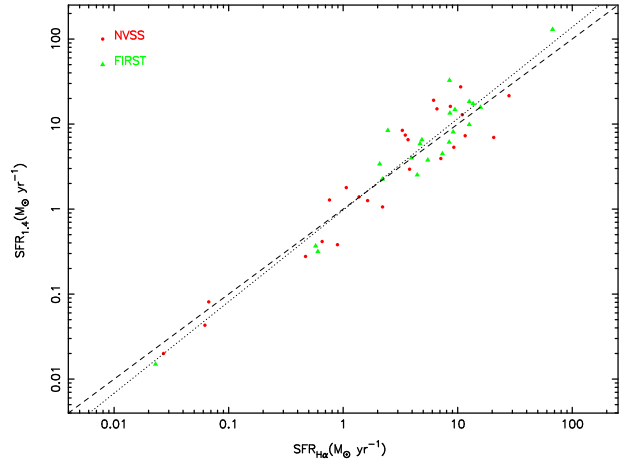}
\caption{Comparison of \sfrha\ with \sfrra\ from both NVSS 
(circles) and FIRST (triangles). The dashed line is the 
relation \sfrha = \sfrra, and the dotted line is a linear 
least-squares fit to the data points. 
The Spearman Rank correlation coefficient between \sfrha\ with 
\sfrra\ is 0.9.
\label{ra-ha}}
\end{figure}

Figure~\ref{ra-ha} shows the comparison between SFRs derived from 
1.4\,GHz (FIRST and NVSS) and the \ha\ luminosities.  If the 
underlying absorption and dust extinction were not corrected for, the 
trend between the \sfrra\ and \sfrha\ is similar to that of \sfrir\ 
and \sfrha\ (see Fig.~\ref{ha-ir}a, b), since the radio 
luminosities and the IR luminosities have a tight correlation. 
We show the estimated SFRs derived from \ha\ 
luminosities, with the underlying stellar absorption and the 
internal dust extinction corrections, as a function of \sfrra\ in 
Figure~\ref{ra-ha} only.  

The Spearman rank correlation coefficient is $r_s=0.9$ for the 
relation between \sfrra\ and \sfrha; this implies that the 
correlation between \sfrra\ and \sfrha\ is significant at the 
$6\sigma$ level. This correlation has been reported by other 
authors for both the case of nearby galaxies (such as Afonso \etal\ 
2003; Hopkins \etal\ 2003) and for more distant galaxies (Flores 
\etal 2004).  Another interesting point about this relation is that 
bright galaxies (high SFRs) tend to have higher \sfrra/\sfrha\ 
ratios than less luminous galaxies. This can be seen clearly from 
the one-to-one line (dashed line) and the ordinary least squares 
fit line (dotted line). For those galaxies with \sfrha $< 1$ 
\,M$_\odot$\,yr$^{-1}$, SFRs derived from 1.4\,GHz radio 
luminosities are less than those derived from \ha\ luminosities; 
the probable reason for this is that the detector in radio band is 
not sensitive, and the uncertainty for those faint galaxy 
observations is large.  The deviation at high radio luminosities, 
as well as the increasing scatter in the correlation, could result 
from a relatively large amount of extinction in those objects 
undergoing the most vigorous star formation, from the absorption 
of Lyman photons by dust, or from an IMF that weights differently 
the high-mass stars mainly responsible for \ha\ and the lower mass 
stars that dominate the supernova numbers (Cram \etal 1998).

In this section, after comparing with \sfroii\ and \sfrha; \sfrir\ 
and \sfrha, we found that subtracting the underlying stellar absorption 
is very important to determine the internal dust extinction, and 
also in estimating the SFRs of galaxies, especially high 
luminosity galaxies. Without the underlying stellar absorption and 
the internal dust extinction correction, the SFRs estimated with 
IR and radio luminosities are larger than those derived from optical 
spectral lines ($\sfrir > \sfrha > \sfroii$). 
This result is consistent with what has been found and shown in 
Madau-type plots in recent years, where the SFRs obtained from UV 
and optical data are much lower than that obtained from mm and submm 
observations at intermediate and high redshifts. In order to reach 
agreement between both determinations, fixed (and somehow arbitrary) 
amounts of extinction have been applied to the UV/optical data, 
because at the moment, the intermediate and high redshift samples 
do not allow a reliable determination of the dust extinction 
(Rosa-Gonz{\' a}lez \etal 2002).
If the underlying Balmer absorption was not subtracted, the dust 
extinction from Balmer lines (such as \ha, \hb) would be 
overestimated. The \sfrha\, and \sfroii\ show a clear excess and the 
excess is much larger for \sfroii\ than for \sfrha ($\sfrir < \sfrha 
< \sfroii$).     
After taking into account the corrections for Balmer absorptions 
and dust extinction, the SFRs in the optical with those in the FIR 
and radio show a very good agreement.

\section{Discussion}

\subsection{Gas depletion timescales}

\begin{figure}
\centering
\includegraphics[angle=-90,width=0.45\textwidth]{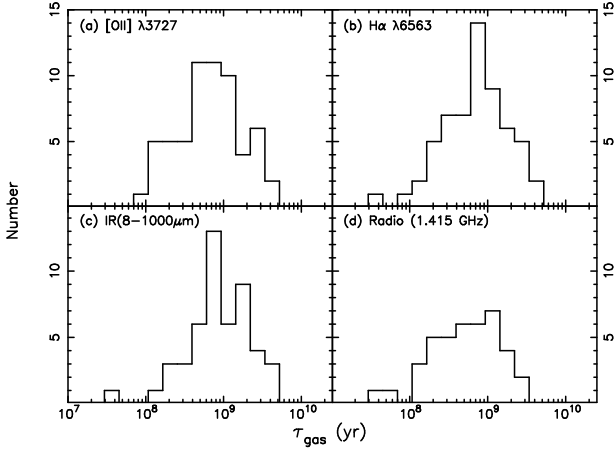}
\caption{The gas depletion time-scale of BCGs from \ha, \oii\ 
emission lines, far-infrared, and radio luminosities. The typical 
gas depletion time-scale of BCGs is about a few billion years.}
\label{tau}
\end{figure}

The star formation history of BCGs is still an open question, BCGs 
are either genuinely young galaxies, or old galaxies that have 
resumed forming stars at a prodigious rate. In the third paper of 
this series (Kong \etal 2003), using a technique of empirical 
population synthesis based on a library of observed star cluster 
spectra, we find that BCGs are typically age-composite stellar 
systems; in addition to young stellar populations, intermediate-age 
and old stellar populations contribute significantly to the 5870 \AA\ 
continuum emission of most 
galaxies in our sample. Since all BCGs have been found to contain 
a population of old stars, our results suggest that BCGs are 
primarily old galaxies with discontinuous star formation histories.

Since the galaxies in our sample were selected from some large 
neutral hydrogen (\hi) surveys (Gordon \& Gottesman 1981; Thuan \& 
Martin 1981), the \hi\ gas mass ($M_{\rm HI}$) of our sample galaxies 
was measured by these \hi\ surveys. The SFR of our sample galaxies 
were measured in Sec. 3, using the \ha, \oii\ emission lines, 
far-infrared, and radio luminosities. Therefore, we can calculate 
the gas depletion time-scale of BCGs, obtained by dividing the 
available gas mass by the present SFR:

\begin{equation}
\tau_{gas}= M_{\rm HI}/{\rm SFR}, 
\end{equation}

where $M_{\rm HI}$ is the total \hi\ mass content and SFR is the 
ongoing star formation rate. 

Figure~\ref{tau} shows the distribution of gas depletion timescales 
for the galaxies in our sample. Most galaxies are consuming their 
interstellar gas at a rate which can only be sustained for another 
few billion years. The median value of the gas depletion time-scale 
of BCGs is less than or about one billion years, and the gas depletion 
time-scale of BCGs is generally much shorter than the age of the 
universe. It indicates that star formation in BCGs could be 
sustained at the current level only on a time scale significantly 
lower than $1/H_0$ (cosmological time) before their neutral gas 
reservoir will be completely depleted.  

To understand the evolutionary connection between blue compact 
star-forming galaxies observed at intermediate redshifts ($0.1 < 
z < 1$) and nearby blue compact galaxies, Pisano \etal(2001) have 
compared Arecibo \hi\ 21 cm spectroscopy and Keck HIRES spectroscopy 
of 11 nearby blue compact galaxies.  Using the \sfrha\ and the \hi\ 
mass, they calculated the gas depletion timescale, $\tau_{gas}= 
M_{\rm HI}/{\rm SFR}$, of each galaxy. The SFRs of BCGs range from 
0.23$-$23.5 M$_\odot$\,yr$^{-1}$, and gas depletion timescales 
range from 0.18$-$7.29 Gyr.  Using the optical spectroscopy, deep 
optical/near-IR photometry of 4 luminous blue compact galaxies, and 
spectral evolutionary models, Bergvall \& {\" O}stlin (2002) 
concluded that star formation in luminous BCGs is intense, 
corresponding to gas depletion timescales of the order of 100 Myr.  
Our results are consistent with those works: the luminous BCGs have 
shorter gas depletion timescale, and the typical gas depletion 
timescale of BCG is about a few billion years.

Based on surface photometry and spectroscopic analysis, most of BCGs 
have been found to contain a population of old stars. Based on the 
gas depletion estimation, the high star formation activities of BCGs 
must be transient. Therefore, the general star formation history 
of BCGs is that these galaxies undergo a few or several short bursts 
of star formation followed by longer more quiescent periods.

\subsection{SFRs versus M$_B$}

\begin{figure}
\centering
\includegraphics[angle=-90,width=0.45\textwidth]{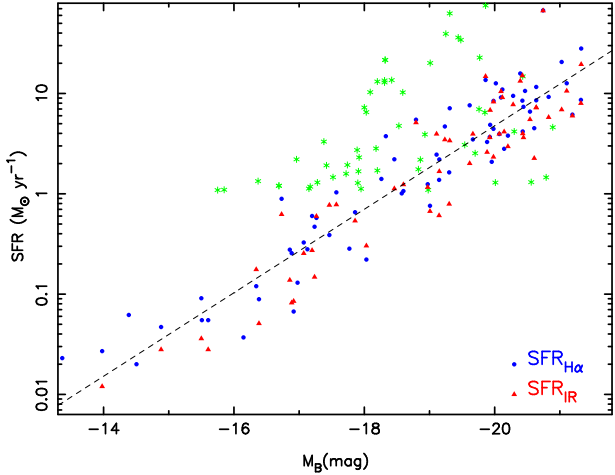}
\caption{ 
Correlation between the star formation rates derived from the 
infrared, \ha\ luminosities and the absolute blue magnitude 
of galaxies. 
The data suggest that there exists a good correlation between star 
formation rates of BCGs and $M_B$. The dashed line shows the 
ordinary least-squares fit for BCGs. 
The different symbols indicate SFRs derived from \ha\ (circles) and 
IR (triangles) for BCGs in this paper, and IR (stars) for galaxies 
from Kong \etal (2004).}
\label{mbsfr}
\end{figure}

Figure~\ref{mbsfr} shows the SFR against absolute B-band magnitude 
for BCGs; the SFRs were derived from the \ha\ (circles) and from IR 
(triangles) luminosities, and $M_B$ were taken from Kong \& Cheng 
(2002a). There is a trend between B-band magnitude and SFR of 
BCGs, the Spearman Rank correlation coefficient being $r_s = -0.9$.  
The higher SFRs are clearly being hosted by galaxies with higher 
luminosities although there is a significant scatter in the trend.  
The least-squares fit of these data, shown by the dash line, yields 
an increased SFR towards high luminosity galaxies, $ SFR = -(7.65 
\pm 0.25) - (0.42 \pm 0.01) {\rm M}_B$. This relation implies that 
the SFR per solar luminosity in BCGs is constant over a range of 7 
magnitudes. 

In Kong \etal (2004), the authors have compiled an  
ultraviolet-selected sample of 115 nearby, non-Seyfert galaxies 
spanning a wide range of star formation activities, from starburst 
to nearly dormant, based on observations with various ultraviolet 
satellites.  To check the correlation between SFR and $M_B$ for 
other galaxy types, we over-plot these galaxies in Figure~\ref{mbsfr} 
as stars.  {\it IRAS} flux densities $f_{\nu} (25\,\micron)$, 
$f_{\nu} (60\,\micron)$ and $f_{\nu}(100\,\micron)$ were used to 
estimate the total far-infrared flux \lir\ and \sfrir. The apparent 
blue magnitudes from De Vaucouleurs \etal (1991) and the redshift 
from the NASA/IPAC Extragalactic Database (NED) were used to calculate 
the absolute B-band magnitudes. From Figure~\ref{mbsfr}, we found 
there is a scattered but good correlation between the $M_B$ and the 
SFR of other galaxy types. Starburst galaxies lie substantially above 
the $M_B \sim {\rm SFR}$ relation for BCGs, and the less vigorously 
star forming galaxies are located in  the region below this relation. Our 
results are consistent with the much large sample of $\sim10^5$ SDSS 
galaxies studied by Brinchmann \etal (2004): they found a clear 
correlation between SFR and stellar mass (luminosity) over a 
significant range in log $M_*$.

While optical spectral line fluxes, UV, IR, and radio luminosity 
are commonly used as SFR indicators, the luminosity at B-band 
wavelengths is also dominated in starburst galaxies by young stellar 
populations. In the absence of other observations, the B-band 
luminosity may thus be used as an SFR indicator for nearby 
large-area image surveys (such as SDSS) and deep image surveys (such 
as COSMOS and  Subaru Deep Field , Ouchi \etal 2003). BCGs can be 
selected according to their compactness and blue colors from images 
in these surveys, and the best-fit relation between ${\rm M}_B$ 
and SFR can be used to estimate the SFR of BCG. 
The SFR based on ${\rm M}_B$ is not very accuracy, since 
the scatter for the ${\rm M}_B$ and SFR relation is large.

\subsection{SFRs and dust extinction}

Star forming galaxies in the local universe have been shown to 
exhibit a correlation between obscuration and FIR luminosity (Wang 
\& Heckman 1996). This implies that the obscuration in a galaxy is 
related to its SFR. This same general effect can also be seen in 
other recent studies (such as Takagi, Arimoto \& Hanami 2003). 
Buat \etal\ (2002) found a good relation 
between the extinction in the \ha\ line and the FIR luminosity for 
starburst galaxies. Hopkins \etal\ (2001) and Sullivan \etal\ (2001) 
report a positive correlation between the dust extinction traced 
by the Balmer decrement and the SFR of the galaxies traced by their 
total FIR or \ha\ luminosities.  

\begin{figure}
\centering
\includegraphics[angle=-90,width=0.45\textwidth]{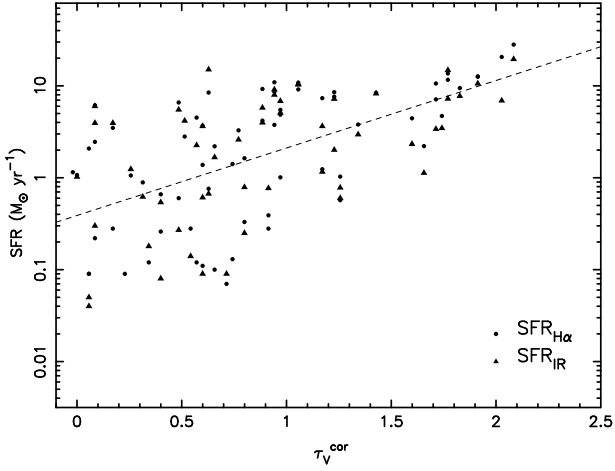}
\caption{The dust extinction traced by the Balmer decrement versus 
the SFR. $\tau_V^{cor}$ were calculated from the \fha\ and \fhb\ 
fluxes where the underlying stellar population absorption were 
subtracted. 
}
\label{sfrebv}
\end{figure}

Figure~\ref{sfrebv} shows the correlation between the dust 
extinction ($\tau_V^{cor}$) and the SFR of BCGs in our sample.  
Using the data of \sfrha\ and \sfrir, we perform a least squares 
fit to the $\tau_V^{cor}$ and \sfr\ to derive $ SFR = -(0.41 \pm 
0.09) + (0.73 \pm 0.0.08) \tau_V^{cor}$.  The Spearman Rank 
correlation coefficient between $\tau_V^{cor}$ and SFR is 0.62. 
There is a clear trend between the dust extinction and the SFR of 
BCGs; the value of SFR increases with $\tau_V^{cor}$. We see 
evidence for a strong correlation between SFR and metallicity (in 
the next subsection), presumably reflecting a correlation between 
metallicity and dust content. The correlation between SFR and dust 
may simply reflect a more fundamental correlation between dust and 
metallicity. This question clearly merits further investigation. 
The scatter in these trends may be related to intrinsic differences 
within galaxy populations. Because greater SFRs correspond to 
greater extinction in the dust, then the SFRs 
of high redshift galaxies derived from either Balmer fluxes or from 
UV continuum measurements would tend to be systematically 
underestimated, if the dust extinction were not corrected. Since 
the dispersion between the dust extinction and the SFR of BCGs is 
large, the situation is worse for these low extinction galaxies 
(most BCGs have low $\tau_V^{cor}$). This relation cannot be 
used to estimate the amount of dust extinction in BCGs.
  
\subsection{SFRs vs. Metallicity}
 
\begin{figure}
\centering
\includegraphics[angle=-90,width=0.45\textwidth]{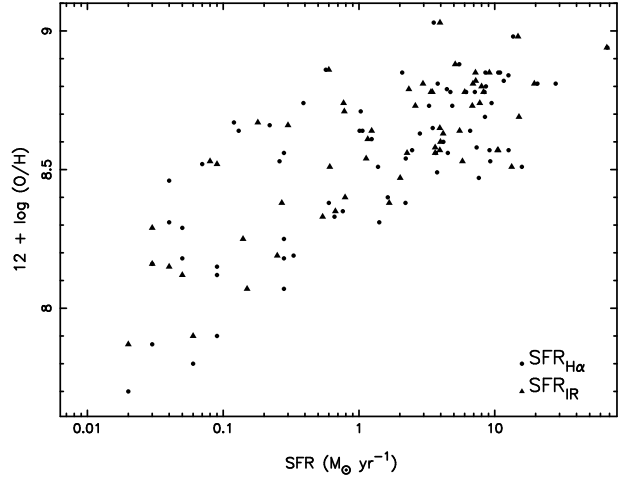}
\caption{Metallicity, 12 + log(O/H), as a function of SFR (\sfrha\ 
and \sfrir). Galaxies with higher global SFRs also show higher 
oxygen abundance.}
\label{sfrz}
\end{figure}

Figure~\ref{sfrz} shows the correlation between oxygen abundance 
and SFR. The oxygen abundance of BCGs was estimated by the N2 = 
\nii$\lambda$6584\AA/H$\alpha$ estimator from the results of 
Denicol{\' o}, Terlevich \& Terlevich (2002). These authors found 
that the N2 estimator follows a linear relation with log(O/H) that 
holds for a wide abundance range, from about $1/50th$ to twice the 
Solar value.  A positive trend of 12+log(O/H) with \sfrha\ and 
\sfrir\ can be seen, in the sense that galaxies with higher global 
SFRs also show higher oxygen abundance. The Spearman Rank 
correlation coefficient for this correlation is 0.67, an unweighted 
linear regression yields $12+{\rm log}(O/H) = 0.24 {\rm log}(SFR) 
+ 8.52$. 

Zaritsky, Kennicutt \& Huchra (1994) found the luminosity 
(mass)-metallicity relationship exists over a large range of 
luminosity, galaxy type, and metallicity. A similar relationship 
was seen for BCGs in our sample (see Fig. 9 in Kong \etal 2002).  
This correlation reflects the fundamental role that galaxy mass 
plays in galactic chemical evolution. With the luminosity and SFR 
correlation (see Figure~\ref{mbsfr}), the most straightforward 
interpretation of the correlation between SFRs and metallicity is 
that more massive galaxies form fractionally more stars in a Hubble 
time (higher SFR) than low-mass galaxies, and then have higher 
metallicity.  From Figure~\ref{sfrz}, we also found that the slope of 
the SFR and metallicity relation for these low-mass galaxies 
increases more quickly and the slope of the more massive galaxies 
increases more slowly. It can probably be explained as in massive 
galaxies, which are already relatively metal rich, it is more 
difficult to significantly alter $12+{\rm log}(O/H)$ than it would 
be in a low-mass, low-metallicity galaxy.

\section{Summary}

In this paper, we have determined the star formation rates and gas 
depletion timescales of 72 blue compact galaxies and investigated 
the source of the discrepancies between these different SFR indicators.  
Then we analyzed the correlation between the star formation 
rate and the absolute B-band magnitude, dust extinction and 
metallicity for the star-forming BCGs in our sample.

Our main results are:
\begin{enumerate}
\item 
The SFRs derived from the \ha, \oii\, emission lines, the {\it IRAS}, 
and the 1.4\,GHz radio luminosities show that the SFR of BCGs cover 
almost four orders of magnitude, from $10^{-2}$ to $10^{2}$ 
M$_\odot$\,yr$^{-1}$.

\item 
The medium and average SFRs from these different indicators for our 
sample galaxies are different. The values derived from IR 
and radio continuum data are much larger than those derived 
from emission lines in the optical spectrum. The sample selection effect 
is important for the different mean SFRs of BCGs in the 
literature, determined using different SFR indicators.  

\item 
Subtracting the underlying stellar absorption is very important when 
calculating both the dust extinction and \sfrha, \sfroii. Otherwise, 
the intrinsic extinction will be overestimated or underestimated; 
star formation rates derived using \oii, \ha, and IR have large 
discrepancies.

\item 
The gas depletion time-scale of BCGs is generally much shorter than 
the age of the universe. Star formation in BCGs could be sustained 
at the current level only on a time scale significantly lower than 
the cosmological time before their neutral gas reservoir will be 
completely depleted.

\item
There are some good correlations between the SFR and the absolute 
B-band magnitude, dust extinction and metallicity for the 
star-forming BCGs. The galaxies with lower luminosity have lower 
metallicity, lower dust extinction and lower SFR.
\end{enumerate}

\begin{acknowledgements}

I thank the anonymous referee, S. Charlot and R. Kennicutt for 
their helpful comments and highly constructive suggestions. I am 
especially grateful to F.Z. Cheng for his valuable discussions 
and spectroscopic co-observation. 
This work is based on observations made with the 2.16m telescope 
of the National Astronomical Observatory of China and supported 
by the Chinese National Natural Science Foundation (CNNSF 
10073009).  
XK acknowledges support provided by the Alexander von Humboldt 
Foundation of Germany and the Japan Society for the Promotion of 
Science (JSPS). 
\end{acknowledgements}

\begin{table}
\caption{SFRs and gas depletion timescales of 72 BCGs.}
\scriptsize
\centering
\begin{tabular}{lccrrrrrrrrrr}
\hline
\hline
Name     & RA(J2000)  &Dec(J2000)  &\multicolumn{4}{c}{SFR ($\Msolar/yr$)}            &M(HI)   &\multicolumn{4}{c}{$\tau_{gas}$ (Gyr)}&$M_B$\\
\cline{4-7}\cline{9-12}
Name     & ($^h$:$^m$:$^s$)&($^{\circ} $:$^{\prime} $ : $^{\prime \prime}$)&[OII]&\ha & IR&1.4GHz               &(\Msolar)  &[OII]&\ha & IR&1.4GHz           &(mag)\\
\hline
iiizw12  &00:47:56.5  &+22:22:23   &   4.913&   3.484&   3.953&   7.413&.300E+10&   0.611&   0.863&   0.760&   0.405& -19.665\\
haro15   &00:48:35.4  &-12:42:59   &   8.188&   6.155&   5.991&  19.061&.554E+10&   0.676&   0.899&   0.925&   0.290& -21.192\\
iiizw33  &01:43:56.5  &+17:03:43   &  14.604&   9.249&   5.789&   5.333&.820E+10&   0.561&   0.887&   1.416&   1.538& -20.827\\
vzw155   &01:57:49.4  &+27:51:56   &  18.089&  20.624&   6.907&   6.957&.000E+00&   0.000&   0.000&   0.000&   0.000& -21.027\\
iiizw42  &02:11:33.5  &+13:55:02   &   7.472&  11.622&   7.253&   7.294&.161E+10&   0.216&   0.139&   0.222&   0.221& -20.637\\
iiizw43  &02:13:45.0  &+04:06:07   &   5.369&   7.127&   3.387&   3.934&.101E+10&   0.188&   0.142&   0.298&   0.257& -19.310\\
iizw23   &04:49:44.4  &+03:20:03   &  26.118&  28.047&  19.512&  21.548&.172E+11&   0.659&   0.614&   0.881&   0.800& -21.327\\
iizw28   &05:01:42.0  &+03:34:28   &   3.984&   2.807&   4.156&   0.000&.000E+00&   0.000&   0.000&   0.000&   0.000& -20.145\\
iizw33   &05:10:48.1  &-02:40:54   &   3.746&   1.635&   0.791&   1.257&.171E+10&   0.456&   1.050&   2.163&   1.358& -19.302\\
iizw40   &05:55:42.8  &+03:23:30   &   0.432&   0.891&   0.624&   0.381&.349E+09&   0.807&   0.392&   0.560&   0.916& -16.731\\
mrk5     &06:42:15.5  &+75:37:32   &   0.072&   0.091&   0.036&   0.000&.130E+09&   1.807&   1.429&   3.614&   0.000& -15.499\\
viizw153 &07:28:12.0  &+72:34:29   &   2.249&   1.375&   0.607&   1.392&.345E+10&   1.535&   2.500&   5.689&   2.483& -19.147\\
viizw156 &07:29:25.4  &+72:07:44   &   3.363&   3.685&   3.645&   6.552&.460E+10&   1.368&   1.250&   1.265&   0.703& -19.925\\
haro1    &07:36:56.4  &+35:14:31   &  11.997&  12.637&  10.607&  18.370&.604E+10&   0.504&   0.480&   0.570&   0.328& -21.105\\
mrk385   &08:03:28.0  &+25:06:10   &   7.768&   8.549&   7.219&  13.437&.105E+10&   0.135&   0.123&   0.146&   0.078& -20.637\\
mrk390   &08:35:33.1  &+30:32:03   &  11.333&   7.361&   3.641&   4.475&.657E+10&   0.581&   0.893&   1.803&   1.469& -20.442\\
zw0855   &08:58:27.4  &+06:19:41   &   4.196&   3.750&   0.000&   0.000&.816E+09&   0.194&   0.218&   0.000&   0.000& -18.330\\
mrk105   &09:20:26.3  &+71:24:16   &   0.746&   1.034&   0.780&   0.000&.000E+00&   0.000&   0.000&   0.000&   0.000& -17.571\\
izw18    &09:34:02.0  &+55:14:28   &   0.022&   0.062&   0.000&   0.043&.905E+08&   4.112&   1.459&   0.000&   2.104& -14.387\\
mrk402   &09:35:19.2  &+30:24:31   &  11.738&   7.626&   2.009&   0.000&.291E+10&   0.249&   0.381&   1.449&   0.000& -19.616\\
haro22   &09:50:11.0  &+28:00:47   &   0.119&   0.089&   0.051&   0.000&.256E+09&   2.153&   2.877&   5.023&   0.000& -16.384\\
haro23   &10:06:18.1  &+28:56:40   &   0.330&   0.256&   0.082&   0.000&.618E+08&   0.187&   0.242&   0.753&   0.000& -16.885\\
iizw44   &10:15:14.7  &+21:06:34   &   3.970&   5.481&   5.124&   3.771&.118E+11&   2.972&   2.153&   2.307&   3.133& -18.792\\
haro2    &10:32:31.9  &+54:24:03   &   3.097&   2.207&   1.127&   1.060&.520E+09&   0.168&   0.236&   0.460&   0.491& -18.461\\
mrk148   &10:35:34.8  &+44:18:57   &   6.994&   4.438&   2.329&   2.522&.196E+10&   0.281&   0.442&   0.841&   0.778& -19.981\\
haro3    &10:45:22.4  &+55:57:37   &   0.731&   0.655&   0.538&   0.415&.563E+09&   0.771&   0.859&   1.047&   1.355& -17.858\\
haro25   &10:48:44.2  &+26:03:12   &  15.048&  15.829&  13.315&  15.759&.574E+10&   0.383&   0.363&   0.432&   0.363& -20.392\\
mrk1267  &10:53:03.9  &+04:37:54   &   1.571&   2.083&   0.000&   3.392&.000E+00&   0.000&   0.000&   0.000&   0.000& -19.955\\
haro4    &11:04:58.5  &+29:08:22   &   0.014&   0.027&   0.012&   0.020&.190E+08&   1.358&   0.703&   1.585&   0.951& -13.978\\
mrk169   &11:26:44.4  &+59:09:20   &   0.588&   0.575&   0.597&   0.368&.522E+09&   0.887&   0.908&   0.875&   1.419& -17.265\\
haro27   &11:40:24.8  &+28:22:26   &   0.496&   0.284&   0.000&   0.000&.231E+09&   0.466&   0.813&   0.000&   0.000& -17.771\\
mrk201   &12:14:09.7  &+54:31:38   &   8.530&  13.625&  14.818&  17.316&.750E+09&   0.088&   0.055&   0.051&   0.043& -19.860\\
haro28   &12:15:46.1  &+48:07:54   &   0.137&   0.067&   0.085&   0.081&.215E+09&   1.570&   3.206&   2.529&   2.655& -16.917\\
haro8    &12:19:09.9  &+03:51:21   &   0.278&   0.277&   0.138&   0.000&.174E+09&   0.627&   0.628&   1.262&   0.000& -16.858\\
haro29   &12:26:16.0  &+48:29:37   &   0.008&   0.023&   0.000&   0.015&.504E+08&   6.295&   2.193&   0.000&   3.357& -13.366\\
mrk213   &12:31:22.2  &+57:57:52   &   2.047&   3.946&   3.954&   3.980&.308E+10&   1.503&   0.780&   0.780&   0.774& -20.071\\
mrk215   &12:32:34.7  &+45:46:04   &  11.041&  12.618&   0.000&   9.867&.000E+00&   0.000&   0.000&   0.000&   0.000& -20.018\\
haro32   &12:43:48.6  &+54:54:02   &   6.237&   4.180&   3.988&   0.000&.882E+10&   1.413&   2.109&   2.208&   0.000& -20.430\\
haro33   &12:44:38.3  &+28:28:19   &   0.090&   0.055&   0.028&   0.000&.198E+09&   2.198&   3.597&   7.079&   0.000& -15.605\\
haro34   &12:45:06.6  &+21:10:10   &   7.219&   8.426&   8.270&   6.124&.704E+10&   0.975&   0.836&   0.851&   1.151& -19.982\\
haro36   &12:46:56.4  &+51:36:46   &   0.016&   0.020&   0.000&   0.000&.137E+09&   8.570&   6.855&   0.000&   0.000& -14.505\\
haro35   &12:47:08.5  &+27:47:35   &   2.537&   2.456&   3.953&   8.415&.258E+10&   1.016&   1.050&   0.653&   0.307& -19.108\\
haro37   &12:48:41.0  &+34:28:39   &   1.157&   1.064&   1.237&   1.788&.981E+09&   0.845&   0.925&   0.791&   0.548& -18.599\\
mrk57    &12:58:37.2  &+27:10:34   &   4.030&   3.287&   2.599&   8.436&.222E+10&   0.551&   0.675&   0.853&   0.263& -19.883\\
mrk235   &13:00:02.1  &+33:26:15   &   3.935&   4.868&   6.817&   6.544&.000E+00&   0.000&   0.000&   0.000&   0.000& -19.930\\
mrk241   &13:06:19.8  &+32:58:25   &   3.102&   4.686&   3.468&   5.874&.195E+10&   0.630&   0.416&   0.562&   0.332& -19.236\\
izw53    &13:13:57.7  &+35:18:55   &   0.309&   0.388&   0.771&   0.000&.100E+10&   3.236&   2.576&   1.297&   0.000& -17.464\\
izw56    &13:20:35.3  &+34:08:22   &  65.419&  67.364&  66.927& 129.683&.000E+00&   0.000&   0.000&   0.000&   0.000& -20.742\\
haro38   &13:35:35.6  &+29:13:01   &   0.080&   0.055&   0.000&   0.000&.884E+08&   1.104&   1.607&   0.000&   0.000& -15.508\\
mrk275   &13:48:40.5  &+31:27:39   &   5.224&   4.505&   2.261&   0.000&.862E+10&   1.652&   1.910&   3.811&   0.000& -20.607\\
haro39   &13:58:23.8  &+25:33:00   &   0.493&   0.327&   0.255&   0.000&.160E+10&   3.243&   4.898&   6.281&   0.000& -17.070\\
haro42   &14:31:09.0  &+27:14:14   &   2.400&   2.200&   1.671&   2.252&.320E+10&   1.334&   1.455&   1.914&   1.422& -19.151\\
haro43   &14:36:08.8  &+28:26:59   &   0.050&   0.037&   0.000&   0.000&.592E+09&  11.830&  15.996&   0.000&   0.000& -16.147\\
haro44   &14:43:24.7  &+28:18:04   &   0.381&   0.280&   0.000&   0.000&.121E+10&   3.177&   4.325&   0.000&   0.000& -17.128\\
iizw70   &14:50:56.5  &+35:34:18   &   0.379&   0.469&   0.148&   0.277&.328E+09&   0.865&   0.700&   2.218&   1.183& -17.236\\
iizw71   &14:51:14.4  &+35:32:31   &   0.207&   0.130&   0.000&   0.000&.921E+09&   4.446&   7.079&   0.000&   0.000& -16.976\\
izw97    &14:54:39.2  &+42:01:26   &   0.214&   0.221&   0.303&   0.000&.000E+00&   0.000&   0.000&   0.000&   0.000& -18.032\\
izw101   &15:03:45.8  &+42:41:59   &   0.982&   1.245&   1.163&   0.000&.392E+10&   3.990&   3.133&   3.381&   0.000& -18.973\\
izw117   &15:35:53.6  &+38:40:37   &   8.874&   9.467&   7.757&  14.817&.819E+10&   0.923&   0.865&   1.054&   0.553& -20.282\\
izw123   &15:37:04.2  &+55:15:48   &   0.040&   0.047&   0.028&   0.000&.575E+08&   1.439&   1.225&   2.056&   0.000& -14.882\\
mrk297   &16:05:12.9  &+20:32:32   &  11.812&   8.483&  15.084&  32.678&.516E+10&   0.438&   0.608&   0.342&   0.158& -20.425\\
izw159   &16:35:21.1  &+52:12:53   &   0.773&   0.601&   0.273&   0.316&.174E+09&   0.225&   0.290&   0.637&   0.551& -17.199\\
izw166   &16:48:24.1  &+48:42:33   &  11.310&   9.131&  10.478&   8.069&.000E+00&   0.000&   0.000&   0.000&   0.000& -20.097\\
mrk893   &17:15:02.2  &+60:12:59   &   1.824&   1.010&   0.000&   0.000&.170E+10&   0.933&   1.683&   0.000&   0.000& -18.574\\
izw191   &17:40:24.8  &+47:43:59   &   3.078&   3.797&   2.962&   2.945&.000E+00&   0.000&   0.000&   0.000&   0.000& -20.202\\
ivzw93   &22:16:07.7  &+22:56:33   &   1.965&   1.407&   0.000&   0.000&.137E+10&   0.695&   0.973&   0.000&   0.000& -18.262\\
mrk303   &22:16:26.8  &+16:28:17   &  11.272&  10.609&   0.000&  27.405&.516E+10&   0.457&   0.486&   0.000&   0.188& -20.465\\
zw2220   &22:23:02.0  &+30:55:29   &   7.713&   8.624&   7.992&  16.154&.535E+10&   0.693&   0.621&   0.670&   0.330& -21.322\\
mrk314   &23:02:59.2  &+16:36:19   &   1.072&   0.759&   0.670&   1.280&.278E+10&   2.600&   3.664&   4.150&   2.173& -19.007\\
ivzw142  &23:20:03.1  &+26:12:58   &   9.980&  10.963&   9.130&  12.901&.256E+10&   0.256&   0.233&   0.281&   0.199& -20.117\\
ivzw149  &23:27:41.2  &+23:35:21   &   9.456&   6.591&   5.509&  15.087&.101E+11&   1.067&   1.531&   1.832&   0.668& -20.541\\
zw2335   &23:37:39.6  &+30:07:47   &   0.138&   0.120&   0.176&   0.000&.173E+09&   1.253&   1.442&   0.984&   0.000& -16.342\\
\hline
\end{tabular}
\label{tab1}
\end{table}
\end{document}